\documentclass[12pt]{iopart}

\usepackage{iopams}

\newcommand{\TCFE}{\mathit{T}_{\mathrm{\scriptscriptstyle C}}^{\mathrm{\scriptscriptstyle FE}}}
\newcommand{\TCL}{\mathit{T}_{\mathrm{\scriptscriptstyle C}}^{\mathrm{\scriptscriptstyle L}}}
\newcommand{\TCH}{\mathit{T}_{\mathrm{\scriptscriptstyle C}}^{\mathrm{\scriptscriptstyle H}}}
\newcommand{\TCFM}{\mathit T_{\mathrm{\scriptscriptstyle C}}^{\mathrm{\scriptscriptstyle FM}}}
\newcommand{\bEex }{\bi E_{\mathrm{ext}}}
\newcommand{\bBex }{\bi B_{\mathrm{ext}}}
\newcommand{\Eex }{\mathit E_{\mathrm{ext}}}
\newcommand{\Bex }{\mathit B_{\mathrm{ext}}}

\newcommand{\Ec }{\mathit E_{c}}

\newcommand{\chiM }{{\chi}_{\mathrm{\scriptscriptstyle M}}}
\newcommand{\hchiFE }{\hat{\chi}^{\mathrm{\scriptscriptstyle FE}}}
\newcommand{\epsSiO }{{\epsilon}^{\mathrm{\scriptscriptstyle SiO_2}}}
\newcommand{\epsFE }{{\epsilon}^{\mathrm{\scriptscriptstyle FE}}}
\newcommand{\EHCl }{\overleftarrow E^{\mathrm{\scriptscriptstyle H}}_{\mathrm{\scriptscriptstyle C}}}
\newcommand{\EHCr }{\overrightarrow E^{\mathrm{\scriptscriptstyle H}}_{\mathrm{\scriptscriptstyle C}}}
\newcommand{\ELCl }{\overleftarrow E^{\mathrm{\scriptscriptstyle L}}_{\mathrm{\scriptscriptstyle C}}}
\newcommand{\ELCr }{\overrightarrow E^{\mathrm{\scriptscriptstyle L}}_{\mathrm{\scriptscriptstyle C}}}
\usepackage{graphicx}
\usepackage{dcolumn}
\usepackage{bm}

\begin{document}


\title{Electric field control of magnetic properties and magneto-transport in composite multiferroics}

\author{O.~G.~Udalov}
\address{Department of Physics and Astronomy, California State University Northridge, Northridge, CA 91330, USA}
\address{Institute for Physics of Microstructures, Russian Academy of Science, Nizhny Novgorod, 603950, Russia}
\ead{oleg.udalov@csun.edu}

\author{N.~M.~Chtchelkatchev}
\address{Department of Physics and Astronomy, California State University Northridge, Northridge, CA 91330, USA}
\address{L.D. Landau Institute for Theoretical Physics, Russian Academy of Sciences,117940 Moscow, Russia}
\address{Department of Theoretical Physics, Moscow Institute of Physics and Technology, 141700 Moscow, Russia}

\author{I.~S.~Beloborodov}
\address{Department of Physics and Astronomy, California State University Northridge, Northridge, CA 91330, USA}

\date{\today}


\begin{abstract}
We study magnetic state and electron transport properties of composite multiferroic system
consisting of a granular ferromagnetic thin film placed above the ferroelectric substrate. Ferroelectricity  and magnetism in this case are coupled by the long-range Coulomb interaction.
We show that magnetic state and magneto-transport strongly depend on temperature, external
electric field, and electric polarization of the substrate. Ferromagnetic order exists at
finite temperature range around ferroelectric Curie point. Outside the region the film is in
the superparamagnetic state. We demonstrate that magnetic phase transition can be driven by
an electric field and magneto-resistance effect has two maxima associated with two magnetic phase
transitions appearing in the vicinity of the ferroelectric phase transition.
We show that positions of these maxima can be shifted by the external electric field and
that the magnitude of the magneto-resistance effect depends on the mutual orientation of
external electric field and polarization of the substrate.
\end{abstract}

\submitto{\JPCM}

\noindent{\it magneto-electric effect, multiferroics, Coulomb blockade, granular}:

\maketitle

\section{Introduction\label{sec:intro}}
Control of magnetic state and magneto-transport properties by an electric field
is one of the main challenges in condensed matter physics and materials science these days~\cite{Ramesh2014, Kimura2014, Herklotz2014,Chien2012,Scott2006,Spal2007,Bar2008,Ohno2008}.
There are several known mechanisms of magneto-electric coupling:
i) electric field influence on a surface magnetic properties of magnetic films~\cite{Givord2007,Oda2009,Tsymbal2008},
ii) spin-orbit interaction of electrons in a single crystal multiferroics~\cite{Bal2005,Ser2006}, and
iii) strain mediated coupling of ferroelectrics (FE) and ferromagnets (FM)~\cite{Nan1994,Schultz2007,Goennenwein2010}.
Also the magneto-electric coupling appearing due to the control of the exchange bias in the multiferroic/ferromagnet interface is actively discussed these days.~\cite{Dynes2013,Fontcuberta2006,Binek2010}
The magnetoresistance can be controlled by electric field using two effects: i) electron spin accumulation
in the vicinity of FE boundary due to screening effects~\cite{Tsymbal2010,Barthelemy2010}, and
ii) electron spin-orbit interaction~\cite{Berakdar2009,Berakdar2011}.
\begin{figure}
\includegraphics[width=0.9\columnwidth]{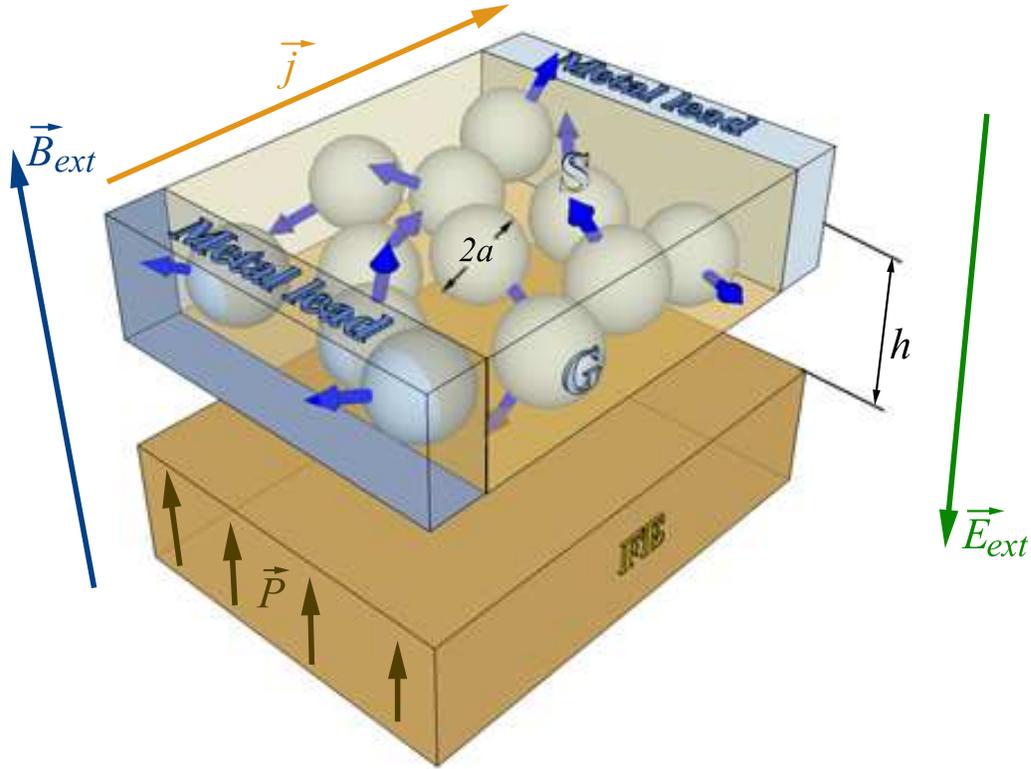}\\
\caption{(Color online) Sketch of composite multiferroic system consisting of a granular ferromagnetic film
placed above the ferroelectric substrate (FE) at distance $h$. Granular ferromagnet consists of ferromagnetic
metallic grains (G) of average size $2a$ embedded in an insulating matrix. Each grain is in the ferromagnetic state with average spin $S$.
Ferroelectric is polarized ($\bi{P}$) perpendicular to its surface. The system is placed into magnetic ($\bi{B}_{\mathrm{ext}}$) and electric ($\bi{E}_{\mathrm{ext}}$) external fields. The electric field is perpendicular to the granular film and does not produce a charge current. Small voltage is applied between the leads leading to the electric current $\bi{j}$.}\label{Fig_1}
\end{figure}

Recently, a new mechanism of magneto-electric coupling was proposed in composite multiferroics --- materials consisting of ferromagnetic grains embedded in a ferroelectric matrix~\cite{Coupl}.  This mechanism is based on the interplay of Coulomb blockade, ferroelectricity, and exchange interaction.  It was shown that the intergrain exchange interaction has a pronounced peak due to strong temperature dependence of dielectric susceptibility of FE component of composite multiferroic in the vicinity of the paraelectric-ferroelectric phase transition. This leads to the unusual magnetic phase diagram of composite multiferroics with FM state appearing in a finite temperature range around FE Curie temperature and superparamagnetic state existing outside this region.

Here we focus on the spatially separated ferroelectric and the granular ferromagnetic films, see figure~\ref{Fig_1}. The long range Coulomb interaction establishes the coupling between the ferroelectric and ferromagnetic degrees of freedom. We show that in contrast to the magneto-electric coupling arising due to the spin-orbit interaction the coupling in composite multiferroics
is non-linear and is similar to the strain mediated magneto-electric effect.

The dielectric permittivity of FE can be controlled not only by temperature but also using an external electric field~\cite{Krupanidhi2006,Waser2006,Yu2004}. This opens the opportunity to control the magnetic state of
the system proximity coupled to FE using the electric field. In this manuscript we predict that electric field can control the
magnetic state of granular ferromagnetic thin film placed above the FE substrate. In particular, we show that electric field can induce the magnetic phase transition in the system. We calculate the magneto-resistance effect and show that it can be controlled by
an external electric field and it depends on the electric polarization of FE component of the system.

The paper is organized as follows. In section~\ref{sec:sys} we introduce the model and discuss important
energy scales. In sextions~\ref{sec:Gap},~\ref{sec:Coul}, and~\ref{sec:Exch} we study the influence of FE
substrate on the intergrain exchange coupling. We investigate magnetic properties of the system
in sections~\ref{sec:Mag} and show that using electric field one can control the magnetization and
magnetic susceptibility. Section~\ref{sec:Phen} describes the magneto-electric coupling
in composite multiferroics using phenomenological theory.
Finally, we study the magneto-resistance (MR) effect as a function of temperature and
electric field.

\section{The model\label{sec:sys}}

We study magnetic and transport properties of the system consisting of magnetic grains embedded in an insulating matrix with FE film (substrate) located in the vicinity of the magnetic granular film, figure~\ref{Fig_1}. The granular film has the
thickness $d$ and is located above the FE substrate at distance $h$, such that $d < h$. The space between the
granular film and the FE substrate is filled with an insulator which is used in granular film.
A similar insulator is located above the granular film.
The magnetic grains have average radius $a$ and the intergrain distance is $r_{\mathrm{g}}$. For simplicity we neglect the distribution 
of grain sizes and the intergrain distance distribution inevitably appearing in granular materials. These effects does not 
play any crucial role for weak magneto-dipole interaction and anisotropy in comparison with the intergrain interaction.

The magnetic granular film is characterized by several energy scales:
1) the magneto-dipole interaction of grains $E_{\rm md}$~\cite{Vinai1999,Trohidou1998}, 2) the magnetic Curie temperature of the material of which grains are made $\TCFM$, 3) the energy of magnetic anisotropy $E_{\mathrm{a}}$ of a single granule which determines  the blocking temperature $T_{\mathrm{b}}$ ($T_{\mathrm{b}}\approx E_{\mathrm{a}}$) at which the fluctuations of grain magnetic moments are suppressed by the anisotropy,~\cite{Livin1959} and 4) the intergrain exchange coupling $J$ and the related ordering temperature~$T_{\mathrm{m}}$~\cite{Boz1972,Hel1981}.

For temperatures $T < \TCFM$ each grain has a finite magnetic moment, typically much larger than $\hbar/2$. For small grains we can neglect the anisotropy energy for temperatures $T > T_{\mathrm{b}}$. We can also neglect the magneto-dipole interaction for temperatures $T\gg E_{\rm md}$. We assume that magnetic state of the system is defined by the exchange interaction. The influence of granular film thickness on the properties of the system is discussed in Appendix~\ref{ApThick}.

The granular film is characterized by the charging (or Coulomb) energy $\Ec=e^2/(a\epsilon)$, which is the electrostatic energy of a single excess electron placed on a grain~\cite{Arie1973,Bel2007review}. We assume that $\Ec\gg T$. In this case the system has an activation conductivity.

The FE substrate is characterized by the polarization $\bi{P}$, which is perpendicular to the substrate surface. The system is placed into external electric $\bEex$ and magnetic $\bBex$ fields.  A small voltage is applied along the granular film leading to the electric current $\bi{j}$.

We note that external electric field can be created by applying voltage between some bottom electrode (below the FE layer)
and the granular film. Since the granular film has a finite conductivity, the electric potential 
will be uniformly spread between all grains. Thus, under applied voltage each grain will have some excess charge.~\cite{Beloborodov2014_ER} This excess charge creates (together with the bottom electrode) a homogeneous electric field governing the 
FE layer. In addition, there is an inhomogeneous part of the electric field generated by grains. 
This inhomogeneous part is responsible for ME coupling.

\section{Coulomb gap\label{sec:Gap}}

The important parameter characterizing the properties of granular film
is the charging (or Coulomb) energy $E_c$. It controls the electron transport~\cite{Arie1973,Bel2007review},
and the exchange coupling between different grains in granular systems~\cite{Coupl}.
This energy depends on the dielectric permittivity of the space surrounding the grain.
Ferroelectrics have tunable dielectric constant. Thus, placing FE in the vicinity of a
grain we can control the charging energy.

Considered system consists of several layers with different electric properties.
The insulating matrix above and below the granular film has the dielectric
permittivity $\epsilon_{\mathrm{\scriptscriptstyle I}}$. The granular film can be treated using the effective medium
approach as an insulator with effective dielectric permittivity $\epsilon_{\mathrm{\scriptscriptstyle G}}$.
We assume that both $\epsilon_{\mathrm{\scriptscriptstyle I}}$ and $\epsilon_{\mathrm{\scriptscriptstyle G}}$ are isotropic and do not
depend on the external electric field. The dielectric permittivity of the
FE layer is anisotropic and depends on the external field. The charging energy of
a metallic sphere placed in such a layered structure can be found numerically,
Appendix~\ref{App:ChargingEn}. Here we use a simplified model which has an
analytical solution. Appendix~\ref{App:ChargingEn} shows that numerical calculations
for more complicated model are in a good agreement with analytical result obtained in this section.

\begin{figure}
\includegraphics[width=1\columnwidth]{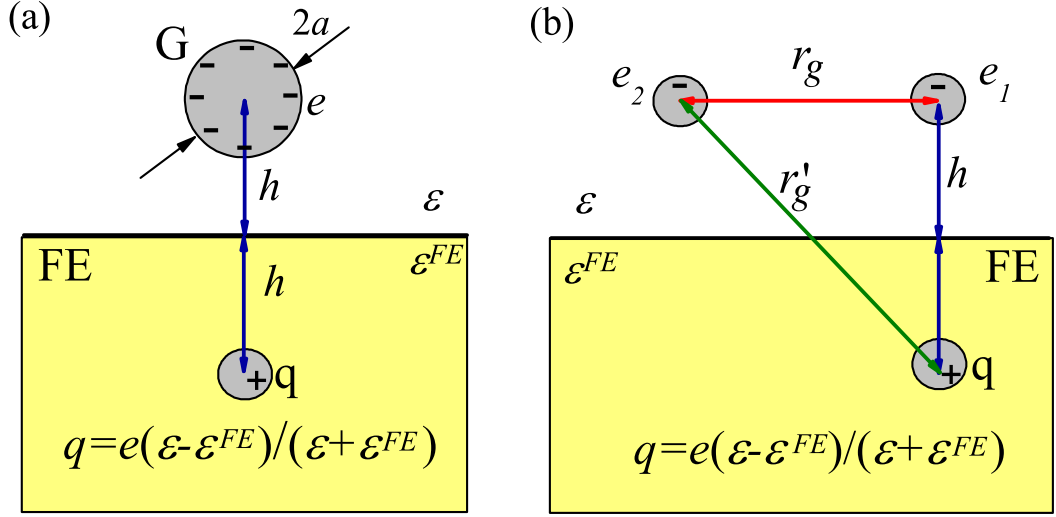}\\
\caption{(Color online) (a) Grain (G) with a single excess
charge $e$ above the FE substrate (FE). Charge image $q$ appears inside the FE substrate.
Interaction with charge $q$ reduces the grain Coulomb energy $E_c$. (b) Two electrons at distance $h$ above the
FE substrate. The distance between electrons is $r_{\mathrm{g}}$, the distance between the
left electron and the charge image $q$ of the right electron is $r^{\prime}_{\mathrm{g}}$.
Interaction of left electron with image $q$ reduces electrons Coulomb interaction $U_c$.
$\epsilon$ and $\epsFE$ are the dielectric constants of the medium above the FE substrate and
the FE substrate, respectively.}\label{ImageCh}
\end{figure}

Consider a metal sphere with radius $a$ placed above the FE at distance $h$, $a < h$, see figure~\ref{ImageCh}(a).
The sphere is charged with the charge $e$. The external electric field $\bEex$ is applied to the
system in the direction perpendicular to the FE surface (the $z$-axis), $\bEex=\Eex \bi z_0$.

The FE substrate is characterized by the position dependent electric polarization
$\bi{P}(\bi{r})=\bi{P}(\bi{E}(\bi{r}))$, where $\bi{r}$ is the position vector. The
electric field $\bi{E}(\bi{r})$ consists of the spatially homogeneous external
field $E_{\rm ext}$ and the inhomogeneous field created by the charged sphere $\bi{E}_{\rm el}(\bi{r})$.
For simplicity we assume that $\Eex \gg |\bi E_{\rm el}|$, and the material relation for FE substrate takes the form $\bi{P}(\bi{E})\approx\bi{P}_0(\Eex)+(\bi{E}_{\mathrm{el}} \cdot \partial_{E_i})\bi{P}_0$.
This expansion is valid for electric field
less than the FE switching field $E_{\mathrm{s}}$.\footnote{If electron is placed at 10 nm above the FE substrate its electrical field is $E_{\mathrm{el}}\approx 0.1$ MV/cm, while the FE switching field, for example for P(VDF/TrFE) FE, is $E_{\mathrm{s}}\approx 1$, thus $E_{\mathrm{el}}\ll E_{\mathrm{s}}$.}
Polarization $\bi{P}_0$ is co-directed with external electric field, $\bi{P}_0=P_0\bi{z}_0$.
The partial derivative defines the dielectric susceptibility tensor of the FE-substrate $\hchiFE=\partial_{E_i}\bi{P}_0$.
We assume that the tensor depends on the polarization $P_0$ and on the external electric field $E_{\rm ext}$. The dielectric tensor of the FE substrate is $\hat{\epsilon}^{\mathrm{\scriptscriptstyle{FE}}}=1+\hchiFE$. Finally, the FE substrate is characterized
by the polarization $P_0(\Eex)$ and the dielectric tensor $\hat{\epsilon}^{\mathrm{\scriptscriptstyle{FE}}}$.

The sphere is located in the media with effective dielectric constant $\epsilon$.
Thus, we replace the layered structure above the FE substrate by the effective medium with
homogeneous electric properties in our simplified model. We discuss the validity of this approach
in Appendix~\ref{App:ChargingEn}. For the Coulomb energy of the charged sphere we find~\cite{landauVol8,Mele2001}
\begin{equation}\label{Gap}
E_c=E_c^{\,0}\frac{1}{\epsilon}\left(1+\frac{a}{h}\frac{\epsilon-\epsFE}{\epsilon+\epsFE}\right),
\end{equation}
where $E_c^{\,0}=e^2/a$ is the Coulomb energy of the charged sphere in vacuum. Following
the paper of Mele ~\cite{Mele2001} we introduce an effective dielectric permittivity of the ferroelectric $\epsFE=\sqrt{\hat{\epsilon}^{\mathrm{\scriptscriptstyle{FE}}}_{zz}\hat{\epsilon}^{\mathrm{\scriptscriptstyle{FE}}}_{xx}}$.
The presence of FE substrate suppresses the Coulomb energy $E_c$ due to the interaction of a charged grain with the image charge appearing inside the FE, figure~\ref{ImageCh}(a). Equation~(\ref{Gap}) does not take into account the work of external field produced
on the charge $e$ assuming that electrons can move in the $(x,y)$-plane only.

\section{Screened Coulomb interaction of two electrons\label{sec:Coul}}

Second important parameter characterizing the properties of granular film
is the screened Coulomb interaction of electrons located in different grains.
We use the same model as in the previous section to find the influence of the
FE substrate on the Coulomb interaction inside the granular film.
The Coulomb interaction in layered system is considered in Appendix~\ref{App:ChargingEn}.
Consider two electrons at distances $h_1$ and $h_2$ above the FE surface and at
distance $l$ apart from each other in the $(x,y)$-plane. Electrons are located in the medium
with effective dielectric constant $\epsilon$ and FE has the permittivity $\epsFE$.
For interaction of two electrons we find~\cite{landauVol8}
\begin{equation}\label{Coulomb}
U_c=U^0\frac{1}{\epsilon}\left(1+\frac{r_{12}}{r^\prime_{12}}\frac{\epsilon-\epsFE}{\epsilon+\epsFE}\right),
\end{equation}
where $r_{12}=\sqrt{l^2+(h_1-h_2)^2}$ is the distance between the electrons, $r^\prime_{12}=\sqrt{l^2+(h_1+h_2)^2}$ is the distance between one of the electron and the image of the second electron inside the FE substrate, and
$U^0=e^2/r_{12}$ is the Coulomb interaction of electrons in vacuum.

\section{Intergrain exchange interaction\label{sec:Exch}}

Here we discuss the intergrain exchange interaction in the system shown in figure ~\ref{Fig_1}.
We assume that all grains are in the FM state and that the grain magnetism is due to the
itinerant (delocalized) electrons. This is valid for transition d-metals
such as Ni, Co and Fe. The magnetic ordering in the whole system appears due to the
intergrain exchange interaction $J$. We assume that exchange interaction between grains is also due to the
intinerant electrons. The wave functions of these electrons located in different grains overlap
leading to the following exchange interaction~\cite{Auerbach},
\begin{equation}
\label{Eq_exch}
J\propto\sum\int\Psi^{*}_1(\bi{r}_2)\Psi_2^{*}(\bi{r}_1)U_c(\bi r_1-\bi r_2)\Psi_1(\bi{r}_1)\Psi_2(\bi{r}_2)\rmd\bi r_1\rmd\bi r_2.
\end{equation}
Here $\Psi_{1,2}$ is the spatial part of the electron wave function located in the
first (second) grain; $U_c$ is the Coulomb interaction of electrons located in different grains.
Summation is over the different electron pairs in the grains.

We assume that the
Coulomb gap, $E_c$, is large and that conduction electrons are localized inside grains.
Outside the grains the electron wave functions exponentially decay
\begin{equation}
\label{WaveFunc}
\Psi_{1,2}(\bi{r}) = A \left\{\begin{array}{l}\rme^{-\frac{a}{\xi}},~~|\bi{r}\pm\bi{r}_{\mathrm{g}}/2|<a, \\\rme^{-\frac{|\bi{r}\pm\bi{r}_{\mathrm{g}}/2|}{\xi}},~~|\bi{r}\pm\bi{r}_{\mathrm{g}}/2|>a. \end{array} \right.
\end{equation}
Here $A = \left(\int|\Psi_{1,2}|^2d\bi r\right)^{-1/2}$ is the normalization constant and $r_{\mathrm{g}}$ is the
distance between two grain centers.
The electron localization length $\xi$ in granular media depends on the Coulomb gap in the
following way $\xi=a/\ln(E^{2}_{c}/T^2g_{\mathrm{t}})$, where $g_{\mathrm{t}}$ is the average intergrain conductance~\cite{Bel2007review}.
It was shown above that the Coulomb gap $E_c$ and the Coulomb intergrain interaction $U_c$
depend on the dielectric constant $\epsFE$ of the FE substrate. Substituting (\ref{Gap}),~(\ref{Coulomb}), and (\ref{WaveFunc}) into (\ref{Eq_exch}) we obtain
the following result for the exchange integral
\begin{equation}
\label{Eq_exch_fin}
J\propto J_0\epsilon^{\frac{4 d }{a}-1}\left(1+\frac{r_{\mathrm{g}}}{r^\prime_{\mathrm{g}}}\frac{\epsilon-\epsFE}{\epsilon+\epsFE}\right)^{-\frac{4 d}{a}}\left(1+\frac{a}{h}\frac{\epsilon-\epsFE}{\epsilon+\epsFE}\right),
\end{equation}
where $r^{\prime}_{\mathrm{g}}=\sqrt{r^2_{\mathrm{g}}+4h^2}$ and $d=r_\mathrm g -2a$. $J_0$ decays exponentially
with increasing the intergrain distance $d$ leading to the decrease of
overall exchange coupling $J$ in Eq.~(\ref{Eq_exch_fin}) with increasing the distance $d$. This is the consequence of the exponential decay of electron wave functions in the insulating matrix. Equation~(\ref{Eq_exch_fin})
shows that the FE substrate strongly influences the intergrain exchange interaction. The large
factor $\frac{4 d}{a}$ in the exchange integral $J$ appears due to the high sensitivity of the exchange interaction
to the ratio of intergrain distance $d$ and the decay length of the wave function $\xi$.
This ratio is defined by the Coulomb blockade effects which can be controlled by temperature and electric field. 
The sensitivity of exchange interaction on dielectric constant $\epsilon$ variations increases with decreasing grain size $a$, since the Coulomb blockade effects become more important in this case. The sensitivity of 
exchange interaction on dielectric constant $\epsilon$  variations increases with increasing 
the intergrain distance $d$. However, the overall exchange interaction decreases in this case. 
Increasing the granular film height $h$ one decreases the coupling since the Coulomb interaction 
between grains and the FE layer decreases in this case.

The
dielectric permittivity of FE, $\epsFE$, can be tuned by the external electric field or temperature,
thus opening the possibility to control the exchange coupling constant.

\section{Magnetic state of granular thin film \label{sec:Mag}}

In this section we study the magnetic state of granular thin film using the mean field
approximation (MFA)~\cite{Vons,Hel1981,Boz1972}. It is valid because the total grain spin is
much larger than $\hbar/2$. We assume that all grains have the same magnetic moment $\mu$ with the total sample magnetization $M = y \mu$, where $y$ is defined
by the following equation
\begin{equation}\label{Lang}
y=\coth\left(\frac{\mu B_{\rm ext}+by}{T}\right)-\frac{T}{\mu B_{\rm ext}+by}.
\end{equation}
Here $\Bex$ is the external magnetic field and $b = b(T)$ is the temperature dependent
Weiss constant. It is related to the magnetic ordering temperature $T_{\mathrm{m}}=b/3$ and
to the microscopic exchange constant $J$, $b=zJ$, where $z$ is
the coordination number~\cite{Vons,Auerbach}. Using (\ref{Eq_exch_fin}) we find
\begin{equation}\label{Weiss}
b=b^0\epsilon^{\frac{4 d}{a}-1}\left(1+\frac{r_{\mathrm{g}}}{r^\prime_{\mathrm{g}}}\frac{\epsilon-\epsFE}{\epsilon+\epsFE}\right)^{-\frac{4 d}{a}}\left(1+\frac{a}{h}\frac{\epsilon-\epsFE}{\epsilon+\epsFE}\right),
\end{equation}
where $b_0$ is the Weiss constant with dielectric constant $\epsilon = 1$.
\begin{figure}
\includegraphics[width=0.9\columnwidth]{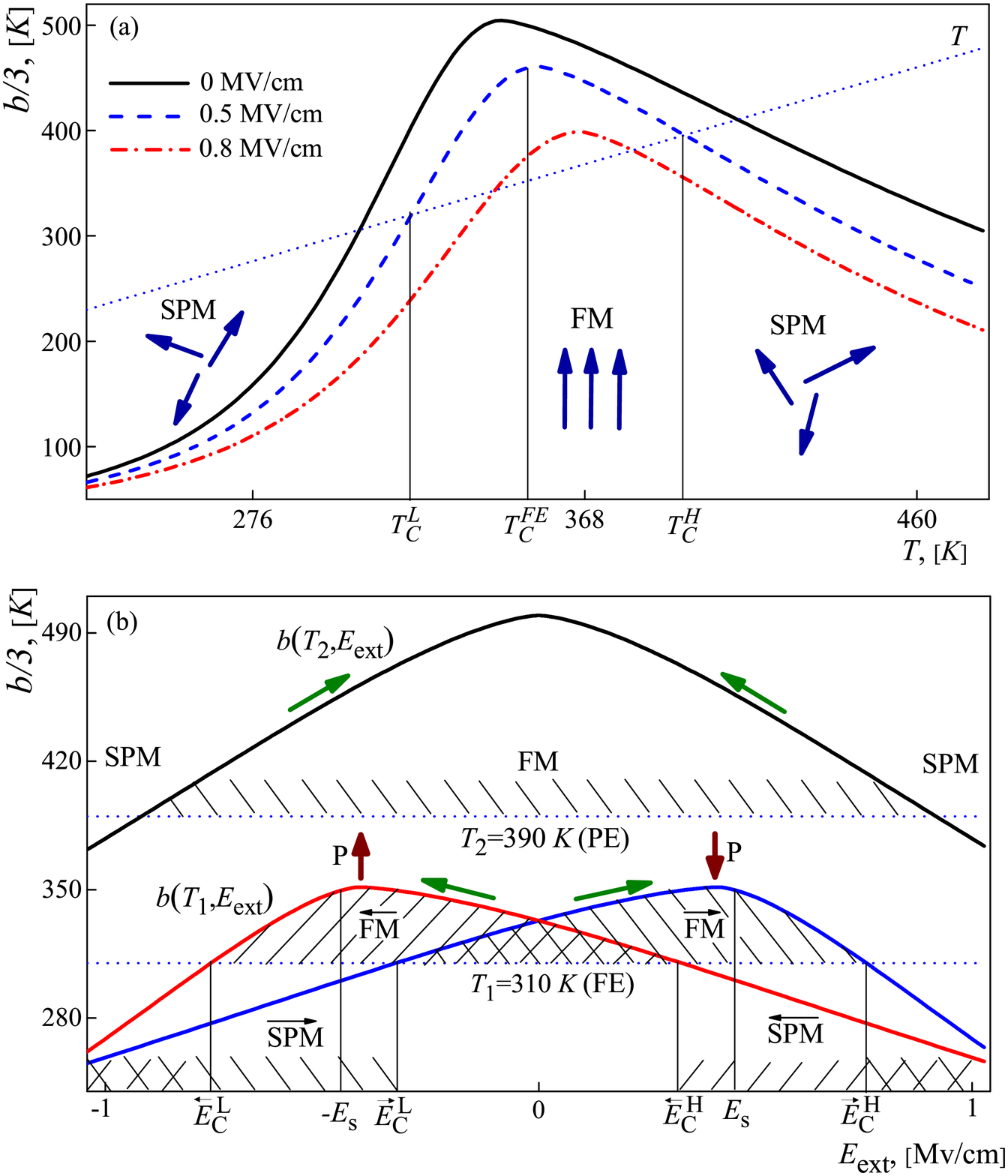}\\
\caption{(Color online) (a) Weiss constant $b$, (\ref{Weiss}), vs. temperature
for different external electric fields $\Eex$. The curves correspond to Ni granular thin film placed above
P(VDF/TrFE) ferroelectric (FE) substrate. Dotted line stands for temperature $T$.
Intersections of temperature and Weiss constant $b$ curves
correspond to the superparamagnetic-ferromagnetic (SPM-FM) phase transition. For temperatures
$\TCL < T < \TCH$ the granular film is in the ferromagnetic (FM) state.
Outside this region the film is in the SPM state. Transition temperatures $\TCL$ and $\TCH$
depend on the external electric field, $\Eex$. (b) Weiss constant $b$ vs.
external electric field $\Eex$ for different temperatures. At high temperature, $T_2=390$K the FE
substrate is in the paraelectric (PE) state with zero spontaneous polarization
of the substrate. At low external electric fields, $b(T,\Eex)>T$, the magnetic moments of granular film
are ordered (FM state). At high electric fields, $b(T,\Eex)<T$, the magnetic film is in
the SPM state. At low temperatures, $T_1=310$K, the FE substrate has finite spontaneous
polarization. The red up and down arrows show the direction of FE substrate polarization $P$.
The green arrows indicate the path around the hysteresis loop.
Critical fields $\EHCl$ and $\EHCr$
stand for electric field induced magnetic phase transitions for one branch and $\ELCl$
and $\ELCr$ for another branch.}\label{bWeiss_T}
\end{figure}

Figure~\ref{bWeiss_T} shows the constant $b$ vs. temperature $T$ and external electric fields $\Eex$ for
granular film consisting of Ni grains with average radius $a=2.5$ nm embedded into SiO$_2$ insulating matrix with
the average intergrain distance $r_g\approx 1.3$ nm and with film effective dielectric constant
$\epsilon_{\mathrm{\scriptscriptstyle G}}=(a+r_{\mathrm{g}})\epsSiO/(2r_{\mathrm{g}})\approx 6$, where $\epsSiO\approx4$ is the dielectric
constant of SiO$_2$. The effective dielectric constant of the upper halfspace in this case is $\epsilon \approx 5$.
The constant $b_0$ is estimated using experimental data of Bozowski~\cite{Boz1972}.
For Ni concentration 0.45\% the FM state appears at temperature $T = 100 K$.

We consider P(VDF/TrFE)(72/28) material for ferroelectric substrate.
It has low dielectric constant ($\epsilon<100$) and well pronounced polarization
hysteresis loop~\cite{Park2007,Ohigashi1983,Kitayama1981}.
We use the notation $\epsFE(T,\Eex)$ for dielectric constant.
The distance between the granular film centre and the ferroelectric
substrate is $h=12$ nm. The behavior of dielectric permittivity vs.
temperature and electric field is shown in Appendix~\ref{ApEps}.

The straight dotted line in figure~\ref{bWeiss_T}(a) stands for temperature $T$. The region
with $b(T)/3 > T$ corresponds to the FM state of the granular film. Since the constant $b$ has a maximum
in the vicinity of FE phase transition point $\TCFE$, the temperature line intersects $b(T)$-curve
twice leading to the existence of FM state in the temperature interval $\TCL < T < \TCH$.

The inverse phase transition occurs at temperature $\TCL$ with magnetic order appearing
with increasing the temperature. Outside the FE region $[\TCL,\TCH]$ the granular film is
in the superparamagnetic state. Figure~\ref{bWeiss_T} shows that the FM state occurs at much
higher temperatures in comparison to granular Ni film without FE substrate~\cite{Boz1972},
where magnetic phase transition appears at $ T = 100 K$.
This is happening due to suppression of the Coulomb blockade
in the vicinity of FE Curie temperature~\cite{Bel2014} and  the increase of
intergrain magnetic coupling. For temperatures $T < 100 K$ the
superparamagnetic-ferromagnetic phase transition appears~\cite{Boz1972}.

The position of FE Curie temperature depends on the external electric field $\Eex$ leading
to the possibility of controlling the magnetic state of granular magnetic film by the external field.
Figure~\ref{bWeiss_T}(b) shows the dependence of the Weiss constant $b$ on the external electric field $\Eex$.
At temperatures $ T > T_2=390$ K the FE substrate is in the paraelectric (PE) state with
dielectric permittivity of the FE substrate monotonically decreasing with increasing
the field $\Eex$. In the PE state the substrate has zero spontaneous polarization.
According to (\ref{Weiss}) the magnetic coupling between grains decreases too.
At low electric fields the granular film is in the FM state and at high
fields it is in the SPM state. Thus, the electric field driven magnetic phase transition
occurs in composite multiferroics.

At low temperatures the FE substrate is in the FE state with finite
polarization $P$ which can be switched by the external field $\Eex$.
The switching field is $\pm E_{\mathrm{s}}$. At these fields the
dielectric permittivity of FE substrate and the intergrain exchange coupling have maxima.
Due to the hysteresis behavior of dielectric constant the magnetic phase transition field depends
on the branch. The fields $\EHCl$ and $\ELCl$ stand for electric field
induced magnetic phase transition for one branch and $\EHCr$ and $\ELCr$ for another branch.
For electric fields $\EHCl < \Eex < \EHCr$ and $\ELCl< \Eex <\ELCr$
the magnetic states are different for different branches. Thus, the magnetic state of granular
film depends on the electric polarization $\bi{P}$.
\begin{figure}
\includegraphics[width=1\columnwidth]{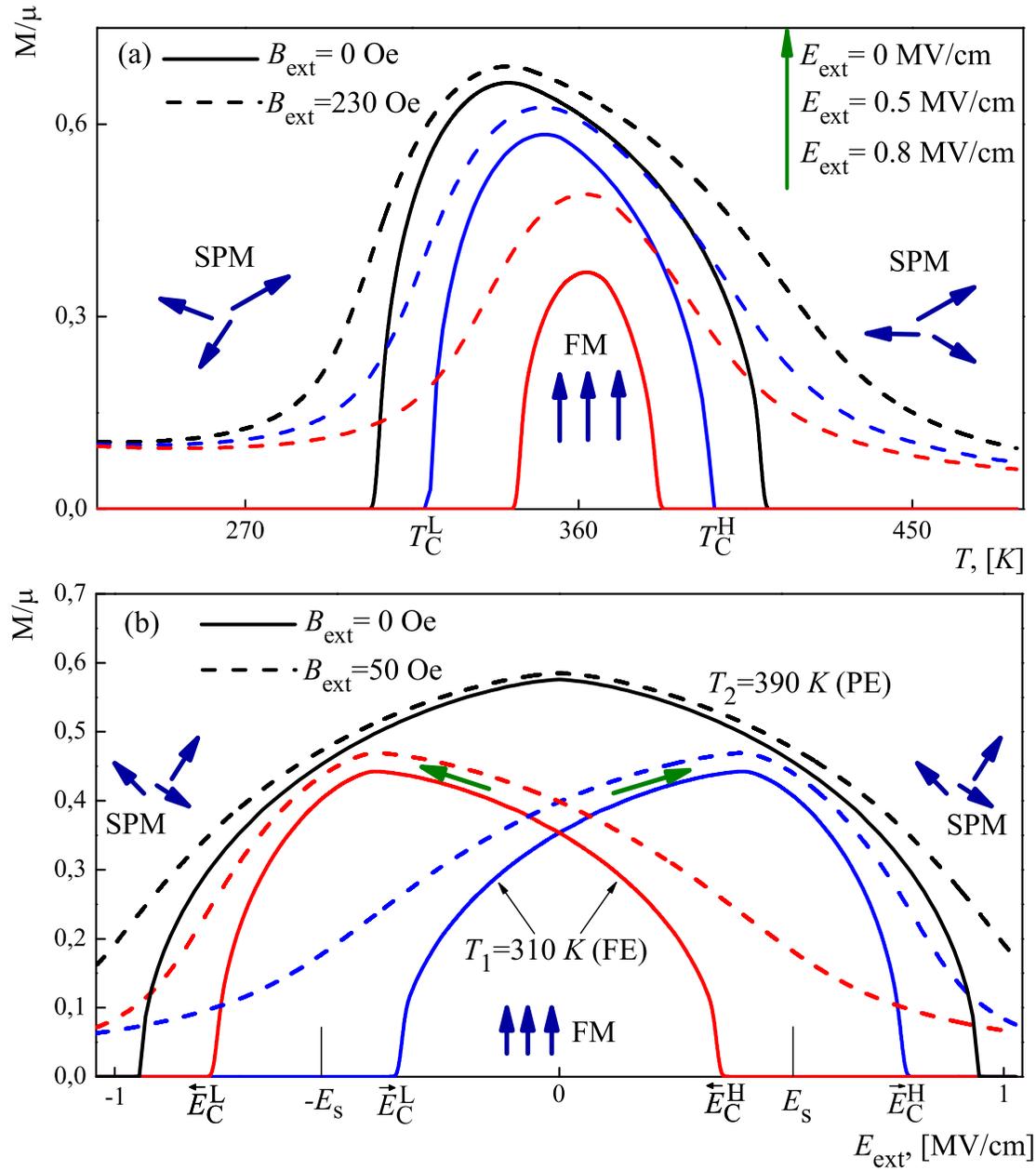}\\
\caption{(Color online) (a) Dimensionless magnetization of granular
film $M/\mu$ in (\ref{Lang}) vs. temperature at different external electric
and magnetic fields. The curves correspond to Ni granular thin film placed above P(VDF/TrFE)
ferroelectric substrate. Regions of ordered (FM) and disordered (SPM)
magnetic states correspond to those in Fig.~\ref{bWeiss_T}(a). Magnetization decreases with
increasing the electric field $\Eex$.
(b) Dimensionless magnetization of granular film $M/\mu$ vs. external electric
field $\Eex$ at different temperatures and external magnetic fields $\Bex$.
Curves correspond to the Weiss constant $b$ shown in figure~\ref{bWeiss_T}(b).}\label{Magn}
\end{figure}

Substituting (\ref{Weiss}) into (\ref{Lang}) we find the magnetization of composite
multiferroics as a function of electric $\Eex$ and magnetic $\Bex$ fields and temperature $T$,
figure~\ref{Magn}. For zero magnetic field, $\Bex = 0$ the average magnetization $M$ is finite
for temperatures $\TCL < T < \TCH$ and it reaches its maximum at the point of
maximum dielectric susceptibility of FE substrate, figure~\ref{Magn}(a).
The position of this maximum depends on the applied external electric field.
For temperatures $T > \TCFE$ the magnetization
monotonically decreases with increasing the electric field up to the point where the
granular film reaches the superparamagnetic state with zero average magnetic moment, figure~\ref{Magn}(b).
Below $\TCFE$ the magnetization is strongly depend on the substrate polarization $P$
due to the hysteresis behavior of FE substrate. For negative polarization
the increase of electric field $\Eex$ ($\Eex < E_{\mathrm{s}}$) leads to the increase of magnetization $M$, figure.~\ref{Epsilon}.
For positive polarization, increasing of electric field destroys the magnetic order.
The external magnetic field $\Bex$ smears the phase transition boundaries.

\begin{figure}
\includegraphics[width=1\columnwidth]{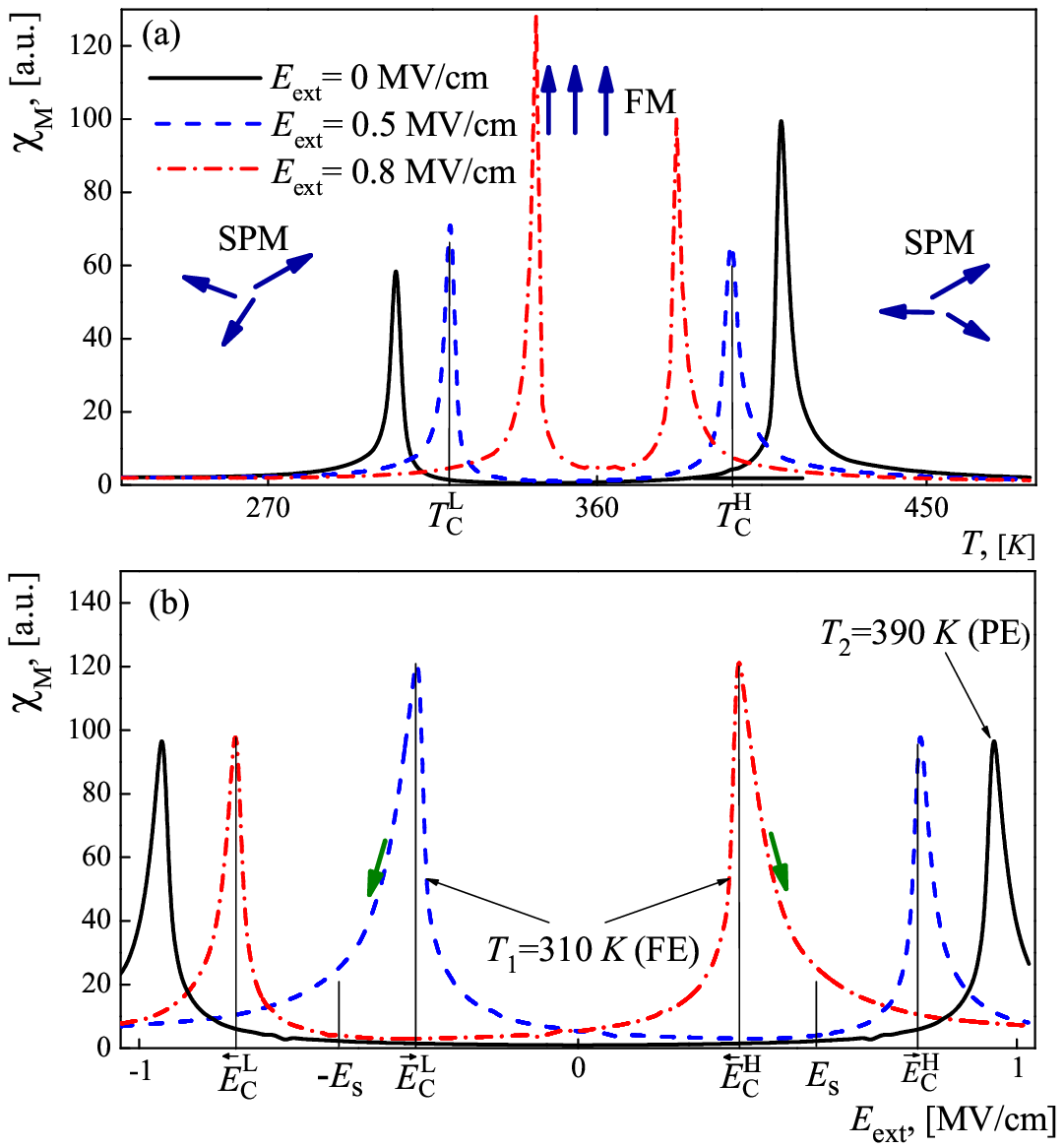}\\
\caption{(Color online) (a) Magnetic susceptibility of granular film $\chiM$ vs. temperature
at different external electric fields. It has peaks in the vicinity of SPM-FM phase
transitions. (b) Magnetic susceptibility $\chiM$ vs. external electric field $\Eex$
at different temperatures. The curves correspond to Ni granular thin film placed
above P(VDF/TrFE) ferroelectric substrate.}\label{Susc}
\end{figure}

Figure~\ref{Susc} shows the magnetic susceptibility of granular
film, $\chiM = \partial M/\partial \Bex$, which can be controlled by the
external electric field and which depends on the electric polarization of the
FE substrate.

\section{Phenomenological description of magneto-electric coupling   \label{sec:Phen}}

Here we discuss a phenomenological description of magneto-electric coupling in
composite multiferroic system consisting of a granular FM film placed above the FE substrate, figure~\ref{Fig_1}.
The Weiss constant $b$ has a quadratic dependence on external electric
field $\Eex$ for small fields and temperatures $T > \TCFE$. For these parameters the
electric polarization of the FE substrate $P$ is linear in field $\Eex$ leading to the quadratic
dependence of the Weiss constant on polarization, $b\sim P^2$. In phenomenological approach the influence
of electric polarization $P$ on the Weiss constant $b$ and on the FM ordering temperature $T_M$ is described
as follows $\gamma^1_{\textrm{me}} M^2 P^2$, where $\gamma^1_{\textrm{me}}$ is some phenomenological
parameter. For temperatures $T < \TCFE$ the Weiss constant $b$
depends on the mutual orientation of polarization and electric field,
$\gamma^2_{\textrm{me}} M^2 (\bi{P}\bi{E}_{\mathrm{ext}})$. Thus, the total ``magneto-electric energy'' is
\begin{equation}\label{phen}
W_{\textrm{me}}=\gamma^1_{\textrm{me}} M^2 P^2+\gamma^2_{\textrm{me}} M^2 (\bi{P}\bi{E}_{\mathrm{ext}}).
\end{equation}
It is important that the magneto-electric coupling in (\ref{phen}) is non-linear, quadratic, in contrast to the
magneto-electric effects appearing due to the spin-orbit interaction~\cite{Scott2006}.

\section{Magneto-resistance of granular multiferroics \label{sec:Tran}}

In this section we consider the magneto-resistance properties of composite multiferroics assuming
that in addition to the external field $\Eex$ there is a small voltage
bias applied along the film producing a finite current.
The conductivity of granular materials is defined by the following
three factors~\cite{Abeles1976,Bel2007review}:
i) the suppression due to Coulomb blockade, $\sim \rme^{-E_c/T}$ with $E_c\gg T$,
ii) the dependence of tunneling conductance on the mutual orientation
of grain magnetic moments, and iii) the influence of exchange interaction on the Coulomb blockade.
The conductivity is given by the following expression,
\begin{equation}\label{Conduct}
\sigma=\sigma^0_{+}\frac{1+\eta}{2}\rme^{-\frac{E_c+E_{\mathrm{m}}}{T}}+\sigma^0_{-}\frac{1-\eta}{2}\rme^{-\frac{E_c-E_{\mathrm{m}}}{T}},
\end{equation}
where $\eta$ is the relative spin polarization of electrons in a single grain,
$\eta=(n^{\uparrow}-n^{\downarrow})/(n^{\uparrow}+n^{\downarrow})$,
with $n^{\uparrow,\downarrow}$ being the numbers of electrons with spins ``up'' and ``down'',
$\sigma^0_{\pm}=\sigma^0(1\pm\zeta \langle M_i\cdot M_j\rangle/\mu^2)$ is the intergrain conductivity
with $\sigma^0$ being the spin independent conductivity, and $\zeta$ being the small parameter describing the
spin dependent tunneling~\cite{Bel2007}.  The quantity
$E_{\mathrm{m}}$ is the average difference of exchange energies inside different grains
\begin{equation}\label{ExchangeDiff}
E_{\mathrm{m}}(T,B_{\rm ext},E_{\rm ext})=\frac{1}{2}J_{\rm int}\cdot\left(1-\frac{\langle M_i\cdot M_j\rangle}{\mu^2}\right),
\end{equation}
with $J_{\rm int}$ being the exchange energy of electrons located in the same grain.
Two energies $J_{\rm int}$ and $J$ are different. The energy $J$ is
the exchange interaction of electrons located in different grains, it defines
the ordering temperature $T_{\mathrm{m}}$; $J_{\rm int}\gg J$. We introduce the notation for
correlation function in (\ref{ExchangeDiff}),
$C_{\mathrm{m}}(T,B_{\mathrm{ext}},E_{\mathrm{ext}})=\langle M_i\cdot M_j\rangle/\mu^2$,~\cite{Abeles1976}.
In MFA it has the form $C_{\mathrm{m}}(T,B_{\mathrm{ext}},E_{\mathrm{ext}})=y^2$.

\begin{figure}
\includegraphics[width=1\columnwidth]{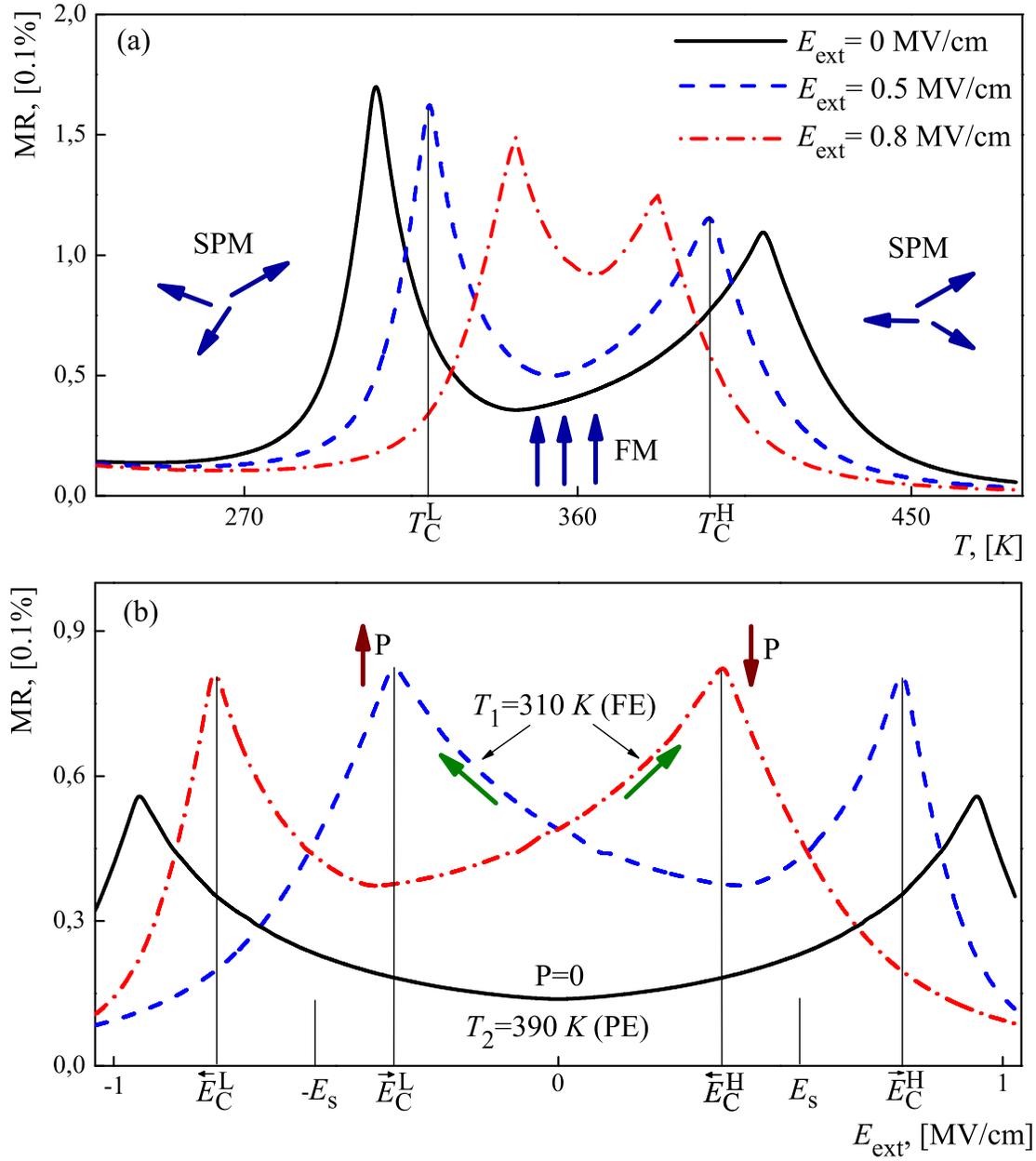}\\
\caption{(Color online) (a) Magneto-resistance MR effect in
composite multiferroic system, (\ref{MR}) and (\ref{MRSimpl}), vs. temperature
at different external electric fields and fixed external magnetic field $\Bex=200$ Oe.
(b) MR effect in granular film vs. external electric field $\Eex$ at different temperatures and
fixed external magnetic fields $\Bex=50$ Oe. All curves correspond to Ni granular thin film placed
above P(VDF/TrFE) ferroelectric substrate.}\label{TMR}
\end{figure}
Using (\ref{Conduct}) and the correlation function $C_{\mathrm{m}}(T,B_{\mathrm{ext}},E_{\mathrm{ext}})$ we find
the magnitude of magneto-resistance effect,
\begin{equation}\label{MR}
\mathrm{MR}(T,\Bex,\Eex)=\frac{\sigma(T,\Bex,\Eex)-\sigma(T,0,\Eex)}{\sigma(T,\Bex,\Eex)}.
\end{equation}
For small electron polarization, $\zeta \ll 1$, $\eta \ll 1$ or small exchange constant $J_{\rm int}$
we find
\begin{equation}\label{MRSimpl}
\mathrm{MR} \approx \eta\left(\zeta+\frac{J_{\rm int}}{2T}\right)\left[C_m(T,\Bex,\Eex)-C_m(T,0,\Eex)\right].
\end{equation}
It follows from (\ref{MRSimpl}) that magneto-resistance effect is controlled by
the expression $\eta\zeta+\eta J_{\rm int}/2T$. Figure~\ref{TMR} shows the MR effect vs. temperature $T$ and
external electric field $\Eex$.

The temperature dependence of MR effect has two pronounced peaks in the
vicinity of PE - FE phase transition in the FE substrate, figure~\ref{TMR}(a).
Each peak is associated with magnetic phase transition~\cite{Boz1972,Hel1981} due to
increase of magnetic fluctuations. In contrast to the materials with single
magnetic Curie point, the composite multiferroics have two magnetic phase transitions
in the vicinity of the FE Curie temperature leading to the occurrence of two MR peaks.

The positions of these peaks and their magnitude is controlled by the
external electric field $\Eex$ applied perpendicular to the system, figure~\ref{Fig_1}.
Figure~\ref{TMR}(b) shows the MR effect vs. external electric field.
For temperatures $T = 390\,\mathrm K > \TCFE$ the FE substrate is in the paraelectric state.
In this case the MR effect has two maxima at certain fields $\Eex=\pm E_c$.
These maxima appear at points of magnetic phase transition driven by the external electric field.

Below the FE phase transition, at temperatures $T=310\,\mathrm K$, the MR effect is more pronounced
since the ratio $\eta J_{\rm int}/2T$ becomes larger. The FE substrate and the MR effect demonstrate
a hysteresis behavior. Arrows in figure~\ref{TMR} indicate the path around the hysteresis
loop. Each branch has two MR maxima associated with magnetic phase transition
driven by the electric field. Therefore the MR effect can be controlled by the
electric field and it depends on the electric polarization of the substrate.

\section{Conclusion}

\begin{figure}
\includegraphics[width=1\columnwidth]{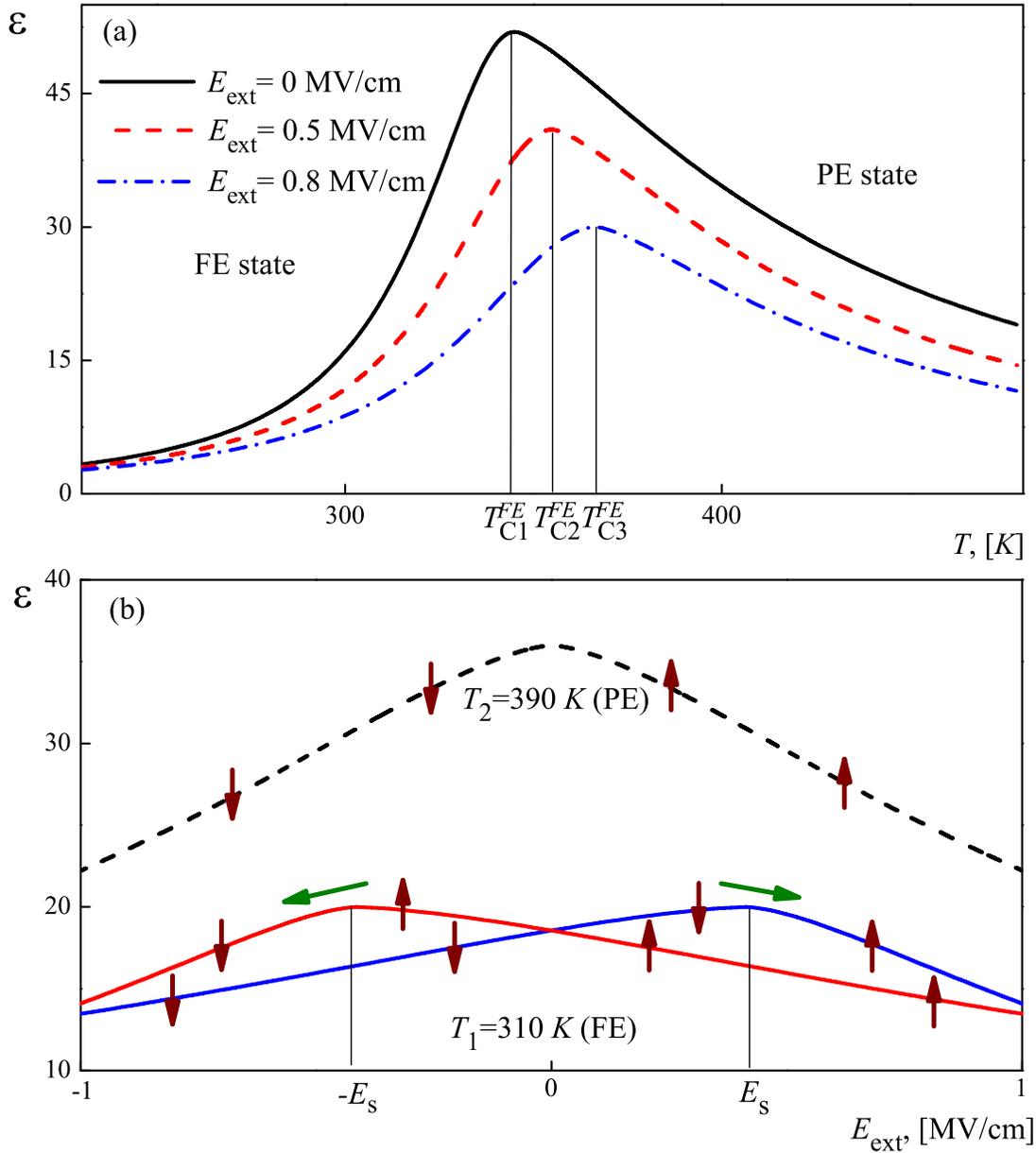}\\
\caption{(Color online) (a) Dielectric permittivity vs. temperature at different
external electric fields $\Eex$. $T_{\scriptscriptstyle{C1},2,3}^{\rm\scriptscriptstyle{FE}}$ stand for temperatures of paraelectric-ferroelectric
(PE - FE) phase transition. (b) Dielectric permittivity vs. external electric field $\Eex$ at different temperatures.
Up and down arrows indicate the direction of average polarization $P$ of the ferroelectric substrate.
At temperature $T=390$~K the polarization $P$ has one component induced by the external electric
field $\Eex$. At $T=310$~K the polarization $P$ has spontaneous and electric field induced components.
Hysteresis loop exists at temperature $T=310$~K. Horizontal arrows indicate the path around
the hysteresis loop. $\pm E_{\mathrm{s}}$ is the switching field of spontaneous polarization.
At these points the dielectric permittivity reaches its maximum.}\label{Epsilon}
\end{figure}

We studied magnetic and transport properties of multiferroic system consisting of
granular ferromagnetic thin film placed above the FE substrate. We showed that
magnetic state of the system strongly depends on temperature, external electric field, and electric
polarization of the FE substrate. The FM state exists at finite temperature range
around the FE phase transition point. Outside this region the superparamagnetic
phase appears. Both the magnetic phase transition temperature and the magnitude of
magnetization are strongly electric field dependent. In addition, the magnetic phase
transition can be controlled by the external electric field. The magnetic state of
the system depends on the mutual orientation of external electric field and polarization
of FE substrate. The ferromagnetic and ferroelectric degrees of freedom are
coupled due to the influence of FE substrate on the screening of intragrain and intergrain Coulomb interaction.

\begin{figure}
\includegraphics[width=0.8\columnwidth]{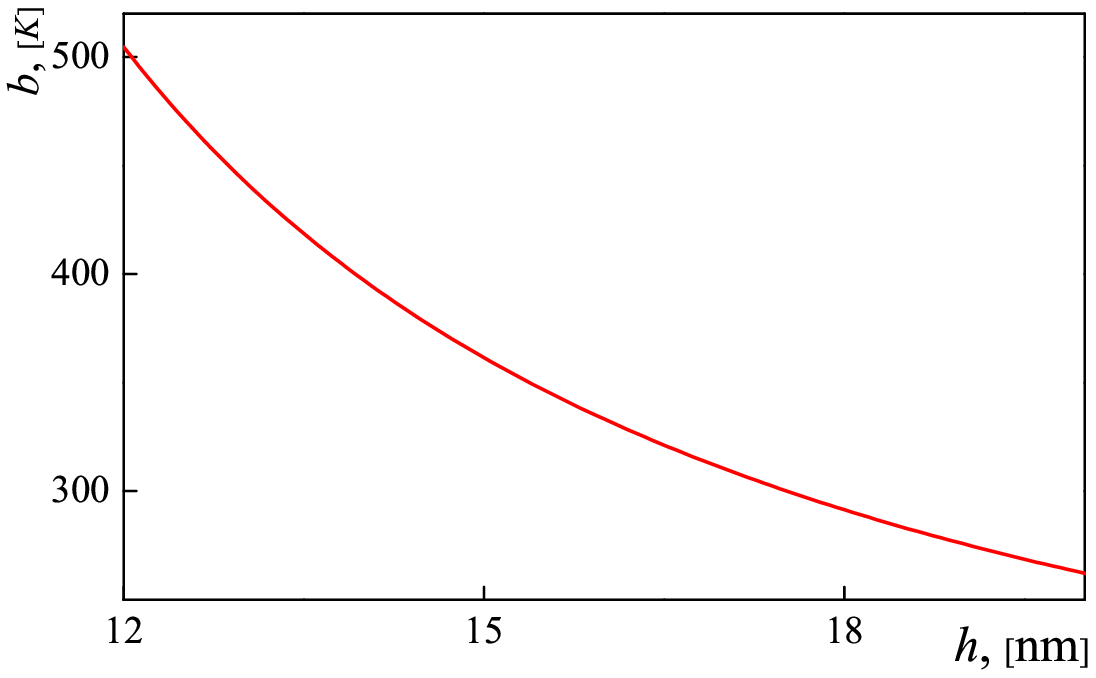}\\
\caption{(Color online) Weiss constant $b$ vs. distance $h$ above the substrate.}\label{Height}
\end{figure}
Also we studied the conductivity of composite multiferroic system and
showed that MR effect strongly depends on temperature, external electric field, and
electric polarization of the FE substrate. The MR effect has two maxima related to two
magnetic phase transitions occurring in the vicinity of FE phase transition. The
positions of these maxima can be shifted by the external electric field. The magnitude of
MR effect depends on the mutual orientation of external electric field and polarization
of the FE substrate.

We demonstrated that magnetic state and MR effect can be controlled
by the electric field for some systems including Ni granular thin film placed
above the FE substrate.

\section{Acknowledgments}

We thank Michael Huth for providing us with his
manuscript prior publication. I.~B. was supported by NSF under Cooperative Agreement Award EEC-1160504 and NSF Award DMR-1158666.
N.~C. was partly supported by RFBR No.~13-02-00579, the Grant of President of Russian
Federation for support of Leading Scientific Schools, RAS presidium and Russian Federal Government programs.

\appendix

\section{Dielectric permittivity vs. temperature and electric field\label{ApEps}}

Here we discuss the behavior of substrate dielectric permittivity as a function of temperature
and external electric field. We estimate the dielectric permittivity of P(VDF/TrFE)(72/28) material
using data of paper \cite{Park2007}.
We use some smooth function to show the important features of $\epsilon(T)$ and $\epsilon(\Eex)$ curves
presented in paper \cite{Park2007}. The dielectric permittivity of P(VDF/TrFE)(72/28) has
temperature hysteresis. In this paper we consider only the ``cooling'' branch for simplicity.
The dependencies $\epsilon(T)$ and $\epsilon(\Eex)$ are shown in figure~\ref{Epsilon}.

\begin{figure}
\includegraphics[width=1\columnwidth]{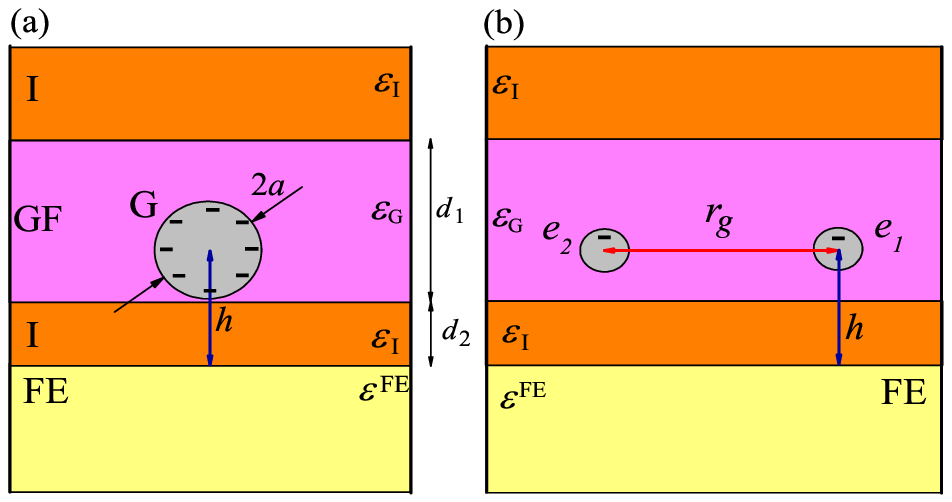}\\
\caption{(Color online) Multilayer system consisting of FE substrate (FE) with
dielectric constant $\epsFE$, insulating layers (I) with dielectric constants $\epsilon_{\mathrm{\scriptscriptstyle I}}$ and
granular film (GF). (a) Metallic grain (G) of size $2a$ is located above the FE substrate at
distance $h$. Space around the grain in granular film is considered as an
effective medium with dielectric constant $\epsilon_{\mathrm{\scriptscriptstyle G}}$. Granular film and the middle
insulating layer have thickness $d_1$ and $d_2$, respectively.
(b) Two charges ($e_{1,2}$) at distance $h$ above the FE substrate and
distance $r_{\textrm{g}}$ apart inside the granular film.}\label{Fig:ML}
\end{figure}

\section{Exchange constant $J$ vs. film thickness\label{ApThick}}

In this appendix we discuss the dependence of exchange constant $J$ on film thickness.
For exchange $J$ in (\ref{Eq_exch_fin}) we assumed
that all grains have the same distance above the FE substrate. This approximation is
valid for thin films with just one layer of grains. For thick granular films the influence of
FE substrate is different for layers located at different distance from the substrate.
However, the dependence of exchange interaction $J$ on grain positions is rather
weak, (\ref{Eq_exch_fin}) and figure~\ref{Height}. Therefore the influence of film thickness is
not important. In addition, the exchange interaction $J$ is exponentially depend
on the intergrain distance $r_{\mathrm{g}}$ leading to a wide distribution of $J$ for different pair of grains.
Averaging over the distance $r_{\mathrm{g}}$ is assumed in our calculations eliminating the dependence of $J$
on the distance from the substrate.

\section{Charging energy of metallic grain in a layered system\label{App:ChargingEn}}

\begin{figure}
\includegraphics[width=0.8\columnwidth]{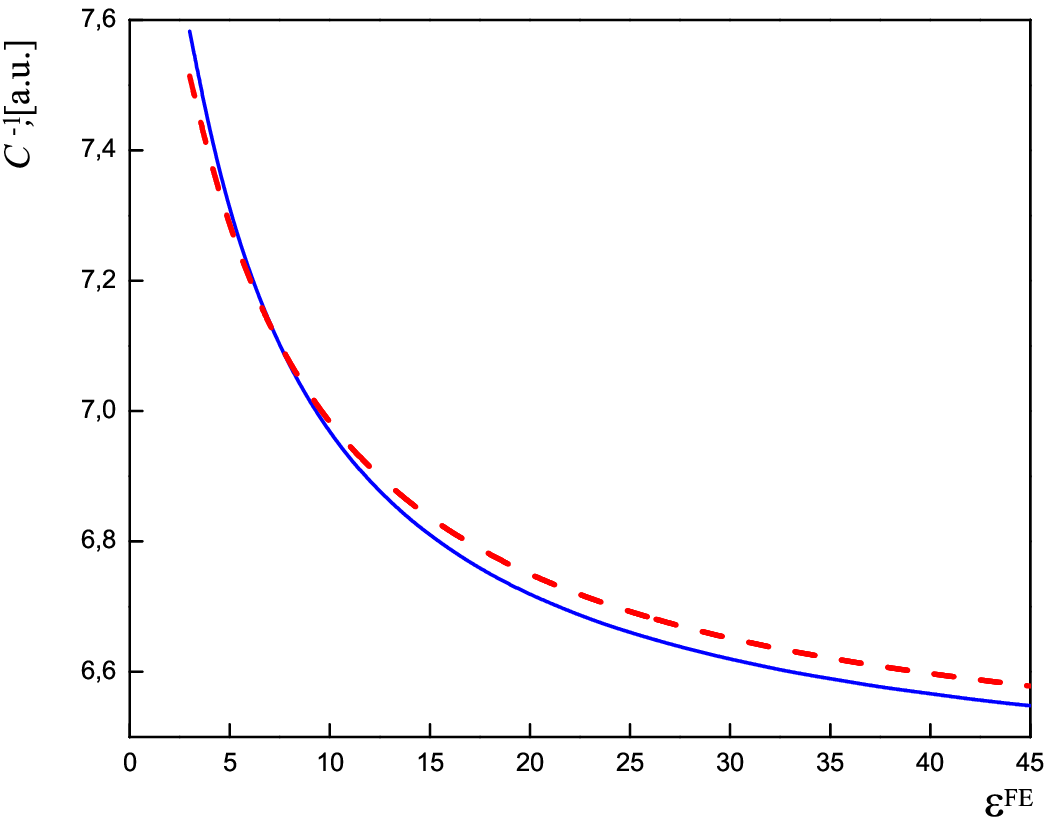}\\
\caption{(Color online) Inverse capacitance $C^{-1}$ of a metallic grain in layered system
shown in figure~\ref{Fig:ML} as a function of dielectric permittivity of FE substrate (perpendicular to the surface component of tensor $\hat{\epsilon}^{\mathrm{\scriptscriptstyle{FE}}}$). Solid line corresponds to the numerical calculations
for the layered system, (\ref{Potential}). Dashed line stands for calculations
with simplified system where upper halfspace being considered as homogeneous with effective
dielectric permittivity $\epsilon$, (\ref{Gap}).}\label{Fig:Cap}
\end{figure}

The charging energy of metallic grain placed in a system consisting of several
insulating layers (see figure~\ref{Fig:ML}) can be estimated as $E_c=e^2/(2C)$ with $C$ being the grain
capacitance. To find the capacitance $C$ we use the source point collocation method.~\cite{Plank2014,Wasshuber2001}
We consider the sphere as the ensemble of point charges $q_i$ placed in the positions $\bi{r}_i$.
We find charges $q_i$ self-consistently assuming that all the points $\bi{r}_i$ have
the same potential $\phi_i=\phi$ and the total charge of the sphere is $Q=\sum q_i$. The potential
at points $\bi{r}_i$ can be found as
\begin{figure}
\includegraphics[width=0.8\columnwidth]{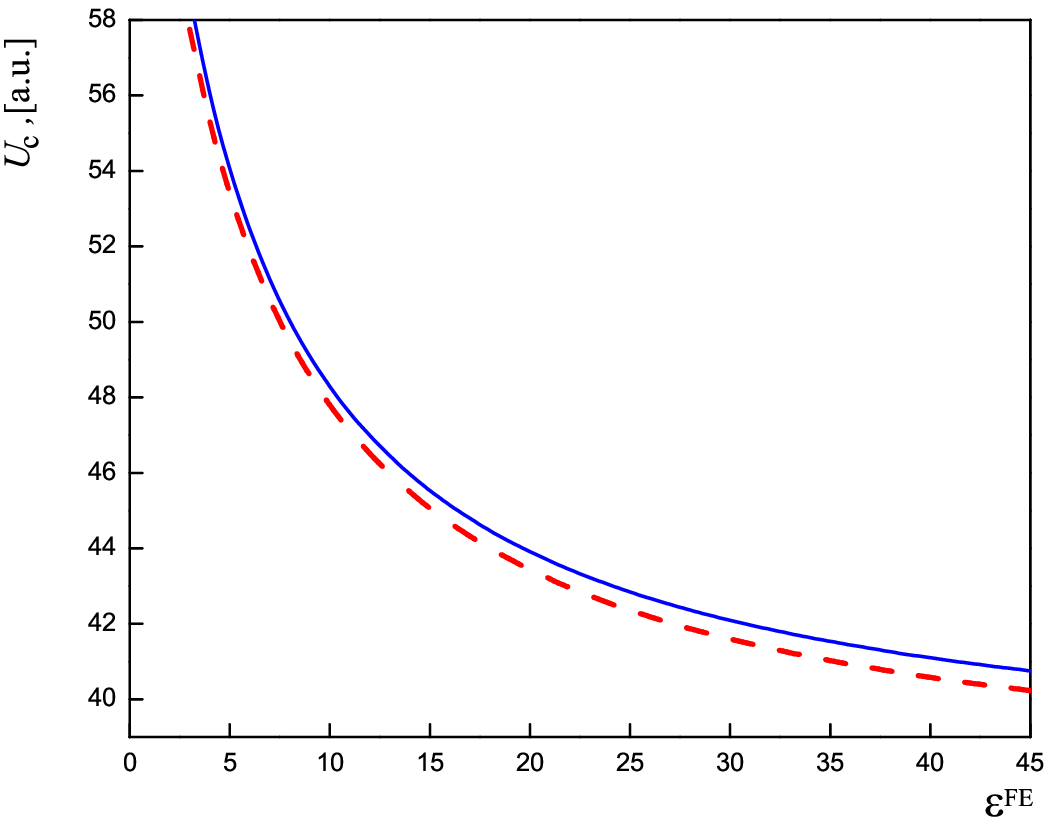}\\
\caption{(Color online) Coulomb interaction $U_c$ of two charges in layered system
shown in figure~\ref{Fig:ML} as a function of dielectric permittivity of the
FE substrate. Solid line corresponds to numerical calculations for the layered system. Dashed line
stands for $U_c$ calculated using the simplified system with upper halfspace being
considered as homogeneous with effective dielectric permittivity $\epsilon$, (\ref{Coulomb}).}\label{Fig:Coulomb}
\end{figure}

\begin{equation}\label{Potential}
\phi_i=\phi=\sum q_j G_{ij},
\end{equation}
where $G_{ij}$ is the electric potential created at point $\bi{r}_i$ by the unit charge located
at point $\bi{r}_j$. The Green functions $G_{ij}$ for layered system can be found
using the two dimensional Fourier transformation. The
capacitance $C=Q/\phi$ can be calculated after solving \ref{Potential}.

We calculate the capacitance $C$ as a function of dielectric permittivity $\epsFE$ of the FE substrate
for the following system: the granular film with $5$ nm Ni grains embedded
into SiO$_2$ matrix is placed above the P(VDF/TrFE) at distance $12$ nm. The
intergrain distance is $1.5$ nm. The thickness of the film is $10$ nm and
the effective dielectric permittivity $\epsilon_{\mathrm{\scriptscriptstyle G}} \approx 5$.
We consider SiO$_2$ insulator with dielectric permittivity $\epsilon_{\mathrm{\scriptscriptstyle I}}=4$
placed above the granular film and in between the FE substrate and the granular film.
The inverse capacitance $C^{-1}$ vs. dielectric permittivity of the FE substrate is shown
in figure~\ref{Fig:Cap}. The dotted line in figure~\ref{Fig:Cap} stands for the inverse
capacitance behavior calculated using (\ref{Gap}) with the effective dielectric
permittivity $\epsilon=4.5$. The numerical calculations with more complicated model
produce almost the same result as the simplified model.

\section{Coulomb interaction of two point charges in a layered system\label{App:Coulomb}}

Here we calculate the Coulomb interaction $U_{c}$ vs. dielectric permittivity of the
FE substrate in layered system (see figure~\ref{Fig:Coulomb}). In figure~\ref{Fig:Coulomb}
the distance between charges is $8$ nm and the distance between charges and the FE is $12$ nm.
The geometrical parameters of the layered system are the same as in the
previous subsection. The dashed line in figure~\ref{Fig:Coulomb} shows the $U_c$
calculated using (\ref{Coulomb}) with $\epsilon=4.5$. The simplified model produces almost
the same result as the more complicated model.

\section*{References}

\bibliography{GFM_coupl}

\end{document}